\documentclass[sn-nature,Numbered]{sn-jnl}

\PassOptionsToPackage{colorlinks=true,linkcolor=black,citecolor=black,filecolor=black,urlcolor=black}{hyperref}

\usepackage{algorithm}%
\usepackage{algorithmicx}%
\usepackage{algpseudocode}%
\usepackage{amsmath,amssymb,amsfonts}%
\usepackage{amsthm}%
\usepackage[title]{appendix}%
\usepackage{booktabs}%
\usepackage{glossaries}
\usepackage{graphicx}%
\usepackage{listings}%
\usepackage{manyfoot}%
\usepackage{mathrsfs}%
\usepackage[version=3]{mhchem}
\usepackage{multirow}%
\usepackage{textcomp}%
\usepackage{units}
\usepackage{xcolor}%
\usepackage{comment}

\setacronymstyle{long-short}
\newacronym{apd}{APD}{avalanche photo-diode}
\newacronym{ase}{ASE}{atomic simulation environment}
\newacronym{df}{DF}{dark field}
\newacronym{dft}{DFT}{density functional theory}
\newacronym{fdtd}{FDTD}{finite-difference time-domain}
\newacronym{pbe}{PBE}{Perdew-Burke-Ernzerhof}
\newacronym{mbe}{MBE}{molecular beam epitaxy}
\newacronym{opa}{OPA}{optical parametric amplification}
\newacronym{shg}{SHG}{second harmonic generation}
\newacronym{sem}{SEM}{scanning electron microscope}
\newacronym{si}{SI}{Supplementary Information}
\newacronym{spdc}{SPDC}{spontaneous parametric down-conversion}
\newacronym{tmd}{TMD}{transition metal dichalcogenide}

\usepackage{xr}

\makeatletter
\newcommand*{\addFileDependency}[1]{
  \typeout{(#1)}
  \@addtofilelist{#1}
  \IfFileExists{#1}{}{\typeout{No file #1.}}
}
\makeatother

\newcommand*{\myexternaldocument}[1]{%
    \externaldocument{#1}%
    \addFileDependency{#1.tex}%
    \addFileDependency{#1.aux}%
}
\myexternaldocument{Main_SI_new}

\begin{document}

\title{Combining ultrahigh index with exceptional nonlinearity in resonant transition metal dichalcogenide nanodisks}





\author*[1]{\fnm{George} \sur{Zograf}}\email{georgii.zograf@chalmers.se}

\author[1]{\fnm{Alexander Yu.} \sur{Polyakov}}

\author[2]{\fnm{Maria} \sur{Bancerek}}

\author[1,2]{\fnm{Tomasz J.} \sur{Antosiewicz}}

\author[1]{\fnm{Bet$\textrm{{\"u}}$l} \sur{K$\textrm{{\"u}}$$\textrm{{\c{c}}}$$\textrm{{\"u}}$k$\textrm{{\"o}}$z}}

\author*[1]{\fnm{Timur O.} \sur{Shegai}}\email{timurs@chalmers.se}


\affil[1]{\orgdiv{Department of Physics}, \orgname{Chalmers University of Technology}, \postcode{412 96}, \city{G$\textrm{{\"o}}$teborg}, \country{Sweden}}

\affil[2]{\orgdiv{Faculty of Physics}, \orgname{University of Warsaw}, \orgaddress{\street{Pasteura 5}, \postcode{02-093}, \city{Warsaw}, \country{Poland}}}


\abstract{Second-order nonlinearity in solids gives rise to a plethora of unique physical phenomena ranging from piezoelectricity and optical rectification to optical parametric amplification, spontaneous parametric down-conversion, and the generation of entangled photon pairs. Monolayer transition metal dichalcogenides (TMDs), such as MoS$_2$, exhibit one of the highest known second-order nonlinear coefficients. However, the monolayer nature of these materials prevents the fabrication of resonant objects exclusively from the material itself, necessitating the use of external structures to achieve optical enhancement of nonlinear processes. Here, we exploit the 3R phase of a molybdenum disulfide multilayer for resonant nonlinear nanophotonics. The lack of inversion symmetry, even in the bulk of the material, provides a combination of a massive second-order susceptibility, an extremely high and anisotropic refractive index in the near-infrared region ($n>$~4.5), and low absorption losses, making 3R-\ce{MoS2} highly attractive for nonlinear nanophotonics. We demonstrate this by fabricating 3R-\ce{MoS2} nanodisks of various radii, which support resonant anapole states, and observing substantial ($>$ 100-fold) enhancement of second-harmonic generation in a single resonant nanodisk compared to an unpatterned flake of the same thickness. The enhancement is maximized at the spectral overlap between the anapole state of the disk and the material resonance of the second-order susceptibility. Our approach unveils a powerful tool for enhancing the entire spectrum of optical second-order nonlinear processes in nanostructured van der Waals materials, thereby paving the way for nonlinear and quantum high-index TMD-nanophotonics.}





\maketitle

\section{Main}\label{sec1}

Nonlinear optics plays a pivotal role in modern science and technology, particularly through second-order nonlinearities that enable crucial processes like second-harmonic generation, optical parametric oscillation, and parametric down-conversion \cite{boyd2020nonlinear,dousse2010ultrabright,lu2019chip}. These nonlinear phenomena require a break of inversion symmetry in the bulk of the material. Consequently, promising material platforms for nonlinear optics applications typically encompass symmetry-broken dielectrics with low optical absorption. Specifically, the degree of second-order nonlinearity is quantified by the second-order nonlinear susceptibility, $\chi^{(2)}$, which gives rise to nonlinear polarization $P^{(2)}_i$ in accordance with:
\begin{equation}
P^{(2)}_i = \sum_{i,j} \varepsilon_0 \chi^{(2)}_{ijk} E_j E_k,  
\end{equation}
where $E_j$ and $E_k$ are the electric field components, and $\varepsilon_0$ is the vacuum permittivity.

\begin{figure}
    \centering
    \includegraphics[width=0.95\linewidth]{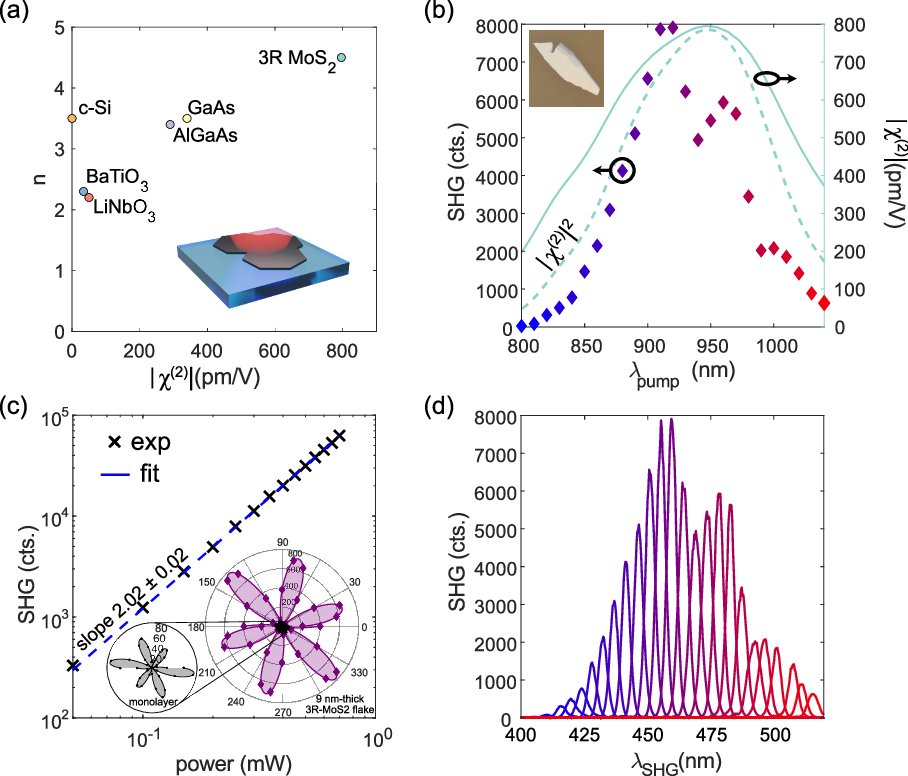}
    \caption{\textbf{3R-\ce{MoS2} material nonlinear optical properties in the 800 -- 1040 nm range.} a) Comparison of linear and nonlinear optical properties for several selected materials -- maximum $|\chi^{(2)}|$ component of each material $vs.$ refractive index $n$ in the near-infrared spectral region. Inset shows the schematic of the optically non-resonant 3R-\ce{MoS2} flake studied in (b-d). Optical data for various materials are taken from: \ce{BaTiO3}~\cite{karvounis2020barium}, $c$-Si~\cite{aspnes1983dielectric,boyd2020nonlinear}, GaAs~\cite{boyd2020nonlinear,shoji1997absolute}, AlGaAs~\cite{koshelev2020subwavelength}, \ce{LiNbO3} from~\cite{shoji1997absolute}, 3R-\ce{MoS2} from~\cite{munkhbat2022optical} for refractive index and the DFT data for nonlinear susceptibility shown in b). b) Colored diamond scatter plot shows a maximum SHG intensity of the optically non-resonant 9~nm thick 3R-\ce{MoS2} flake on a glass substrate as a function of the pump wavelength $\lambda_{\mathrm{pump}}$ at 0.3~mW average incident power for every pump wavelength. Inset shows the optical image of the flake. Cyan solid line shows $|\chi^{(2)}|$ calculated using DFT. Dashed line corresponds to normalized $|\chi^{(2)}|^2$. c) Power-to-power log-log scale plot of SHG intensity $vs.$ the pump power. Inset shows polar plots of polarization-resolved SHG obtained at a 910~nm pump for a 9~nm thick 3R-\ce{MoS2} flake (purple) and \ce{MoS2} monolayer (black), respectively. d) SHG spectra of the R-\ce{MoS2} flake for different pump wavelengths $\lambda_{\mathrm{pump}}$ from 800~nm to 1040~nm with 0.3~mW power. $\lambda_{\mathrm{SHG}}$ corresponds to $\lambda_{\mathrm{pump}}/2$. The color code of the lines corresponds to one of the diamonds in b).}
    \label{fig:1}
\end{figure}

\autoref{fig:1}a illustrates a comparative analysis of commonly used nonlinear materials alongside crystalline silicon ($c$-Si). The latter, although widely regarded as a prominent nanophotonics material due to its high refractive index $n \sim 3.5$ in the lossless near-infrared region \cite{aspnes1983dielectric}, lacks $\chi^{(2)}$ due to its centrosymmetry \cite{boyd2020nonlinear}. Lithium niobate (\ce{LiNbO3}, LN) and barium titanate (\ce{BaTiO3}, BTO) are among the most extensively employed nonlinear materials renowned for their performance in second-order nonlinear processes and integrated photonics applications \cite{carletti2019second,solntsev2018linbo3,fedotova2020second,okoth2019microscale,santiago2021photon,pohl2020integrated,karvounis2020barium,timpu2017second,savo2020broadband}. The advantage of these wide-band semiconductors is their exceptionally low material losses. In the near-infrared region, \ce{LiNbO3} and \ce{BaTiO3} support moderate $\chi^{(2)}$ (largest components) of 50 pm/V \cite{shoji1997absolute} and 34 pm/V, respectively \cite{karvounis2020barium}. A significant weakness of these two materials is the relatively low refractive index in the near-infrared, which is slightly greater than 2, making the optical contrast between the material and the substrate dramatically low. Such low contrast, although allowing for them to perform well in low-loss waveguides and high-quality factor resonators (such as whispering gallery modes, microrings, etc.) for integrated photonics applications and discrete optical components \cite{fedotova2022lithium,zhang2019broadband}, makes it difficult to create compact subwavelength resonant nanostructures supporting optical nonlinear effects.

Having a refractive index similar to that of $c$-Si, both GaAs and AlGaAs benefit from a significant value of $\chi^{(2)}$, of the order of 340 pm/V \cite{shoji1997absolute} and 290 pm/V \cite{koshelev2020subwavelength}, respectively (\autoref{fig:1}a). These materials have already demonstrated outstanding performance with record-high second-harmonic generation (SHG) \cite{koshelev2020subwavelength, timofeeva2018anapoles,frizyuk2019second} and spontaneous parametric down-conversion (SPDC) \cite{marino2019spontaneous,santiago2022resonant}. However, one of the major disadvantages of using GaAs/AlGaAs is the complexity and high cost of fabrication, which requires molecular-beam epitaxy (MBE) techniques. Furthermore, the fabrication process can be potentially intricate due to the requirement of high-index substrates during the growth phase, followed by the transfer to low-index substrates~\cite{koshelev2020subwavelength,camacho2016nonlinear}.

TMDs have recently emerged as a promising high-index nanophotonics platform in the near-infrared range, with a significant second-order nonlinear susceptibility in their atomically thin structures \cite{kumar2013second,li2013probing,wang2015giant,trovatello2021optical,khan2022optical,kumar2022light}. However, their bulk counterparts are unlikely to exhibit substantial $\chi^{(2)}$ due to the commonly adopted A$_1$B$_2$ stacking configuration. This is evident from a vanishing SHG from 2H-MoS$_2$ and similar materials multilayers containing an even number of layers in the few-layer regime and in the bulk of the material~\cite{kumar2013second,li2013probing}. Nevertheless, the utilization of nanophotonic resonances in centrosymmetric multilayer TMD nanostructures has shown a remarkable enhancement in SHG \cite{busschaert2020transition,nauman2021tunable,popkova2022nonlinear}. Furthermore, the SHG intensity of TMD monolayers placed on external metasurfaces has been significantly boosted \cite{hu2019coherent,bernhardt2020quasi}. However, this approach does not allow for implementing an all-TMD nonlinear nanophotonics concept \cite{verre2019transition,munkhbat2020transition,ling2021all,sung2022room,munkhbat2023nanostructured,zotev2023van}. As a result, there is a growing demand for alternative material platforms that can combine high refractive indices with exceptional nonlinearities.

Unlike the 2H counterpart, the 3R phase of \ce{MoS2} does not restore inversion symmetry, allowing it to possess second-order nonlinearity even in the bulk crystal. Recently it was shown that 3R-\ce{MoS2} has one of the most significant $\chi^{(2)}$ values, which is useful for SHG \cite{xu2022towards} and piezoelectricity \cite{dong2023giant} applications. Furthermore, the low optical absorption in the near-infrared range \cite{munkhbat2022optical} makes 3R-\ce{MoS2} advantageous for nonlinear optics applications compared to lossy Weyl semimetals with giant $\chi^{(2)}$ values \cite{wu2017giant}. Additionally, 3R-\ce{MoS2} supports ultrahigh refractive index ($n>4.5$)~\cite{munkhbat2022optical} and has prospects for epitaxial wafer-scale fabrication~ \cite{li2021epitaxial,liu2022uniform}. Therefore, 3R-\ce{MoS2} is an excellent candidate for nonlinear all-dielectric nanophotonics (\autoref{fig:1}a).

Our density functional theory (DFT) calculations predict that $\chi^{(2)}$ of 3R-\ce{MoS2} is particularly high. Specifically, solid lines in \autoref{fig:1}b correspond to several selected $\chi^{(2)}_{ijk}$ tensor components (see right $y$-axis). The cyan-colored line depicts the most dominant ones -- $\chi^{(2)}_{xyx}, \chi^{(2)}_{yyy}, \chi^{(2)}_{yxx}$, which have pronounced resonant behavior peaking at $\lambda_\mathrm{pump}$ of $\sim$ 950~nm and reaching values as high as $\sim$ 800~pm/V. Therefore, 3R-\ce{MoS2} exhibits superior performance compared to conventional nonlinear materials, as summarized in \autoref{fig:1}a, in terms of both second-order nonlinearity and refractive index. The measured SHG as a function of the pump wavelength for a 9~nm thin, optically non-resonant flake qualitatively agrees with the DFT calculations and peaks around 910 -- 920~nm. Moreover, an even better agreement is met when the experimental SHG intensity is plotted against $|\chi^{(2)}_{xyx}|^2$, since $I_\mathrm{SHG} \propto |\chi^{(2)}|^2$. We attribute the observed bell-shaped SHG spectrum to a material resonance of nonlinearity in 3R-\ce{MoS2}. This resonance is likely of excitonic origin, similar to previously observed SHG in monolayer \ce{MoS2} \cite{kumar2013second} and \ce{WSe2} \cite{wang2015giant}. 

To verify the second-order nature of the observed nonlinear process, we plot the power-to-power dependence of SHG at the optimal 910~nm pump. \autoref{fig:1}c shows the incident pump power versus SHG intensity in a log-log scale. The slope of the fitting function is $2.02 \pm 0.02$ with a 95\% confidence, manifesting the second-order nonlinear process. The damage threshold for this flake occurs at laser powers between 50 and 72~mW,
which is in agreement with previous reports \cite{xu2022towards}. Note that the laser used in this experiment has a repetition rate of 80~MHz and a pulse duration of $\sim$ 100~fs, implying that an averaged power of 50~mW corresponds to a peak power of $\sim$ 6.25$\times$10$^3$~W. The latter is nearly 500 times higher than the damage threshold of AlGaAs nanodisks on \ce{SiO2}/ITO~\cite{koshelev2020subwavelength}, making 3R-\ce{MoS2} attractive for high-power nonlinear optics applications.

Furthermore, we performed polarization-resolved SHG measurements on the same flake (optical image is shown as an inset in Figure~\ref{fig:1}b). The result displays a 6-fold symmetry profile, as shown in \autoref{fig:1}c by a purple polar plot. For comparison, we also provide the SHG data of a \ce{MoS2} monolayer, shown in black. The monolayer signal is more than 10 times weaker than that of the 9~nm thick 3R-\ce{MoS2} flake measured under the same conditions.
Additionally, we measured the SHG spectra as a function of excitation wavelength in the 800 -- 1040~nm range at a fixed pump power to demonstrate the precise correspondence between the pump and the emission wavelength, namely $\lambda_{\mathrm{SHG}} = \lambda_{\mathrm{pump}}/2$, \autoref{fig:1}d. The color codes used in \autoref{fig:1}d are the same as in \autoref{fig:1}b. Remarkably, the SHG spectra are intense and lack any residual or background signal, allowing us to use an alternative SHG detection scheme -- an avalanche photodiode (APD) -- which is advantageous for extended SHG mapping experiments discussed below.

\textbf{Optically-resonant nanostructures.} We successfully optimized the optical pump conditions to achieve the maximum SHG intensity in optically non-resonant 3R-\ce{MoS2} flakes, relying solely on the resonant properties of the material. In addition, the DFT calculations were performed using the bulk model of 3R-\ce{MoS2}, which excluded any consideration of optical resonances. Our current objective is to go beyond this and engineer the resonant optical response, in combination with the material properties, to further enhance the efficiency of SHG. The core concept of this work is to ensure that the engineered resonant optical response of the nanostructures aligns spectrally with the material resonances of the 3R-\ce{MoS2}, creating an effective overlap between them. Indeed, the SHG can be dramatically enhanced by using high-$Q$ optical resonators. If the resonance is tuned to the pump wavelength, the SHG output power scales as $P^{2 \omega}\propto\left[Q_1 P^\omega\right]^2$ \cite{koshelev2020subwavelength,frizyuk2019second,popkova2022nonlinear}, where $P^{\omega}$ is the pump power, $\omega$ is the pump photon frequency, and $Q_1$ -- the quality factor of the resonance at the pump wavelength. Here, we do not account for the optical mode at the SHG wavelength due to the high optical losses of 3R-\ce{MoS2} in that range.

We proceed with the design and fabrication of 3R-\ce{MoS2} subwavelength optical resonators. Due to challenges in fabricating spherical nanoresonators capable of hosting optical Mie modes \cite{tselikov2022transition,chernikov2023tunable}, TMD nanodisks are preferred for their highest possible symmetry \cite{verre2019transition,green2020optical}. Specifically, TMD nanodisks are insensitive to incident linear polarization and can support optical modes across visible and near-infrared ranges. Therefore, in this work, we focus on fabricating and studying optically resonant subwavelength nanodisks. These nanodisks can support anapole states in the 880 -- 960 nm region, which enables efficient spectral overlap with the material-specific $\chi^{(2)}$ resonance. The anapole states can substantially enhance the SHG signal by capturing optical excitations within the disk instead of radiating them out to the far field \cite{busschaert2020transition,baranov2018anapole,timofeeva2018anapoles,ushkov2022anapole}.

\begin{figure}
    \centering
    \includegraphics[width=0.95\linewidth]{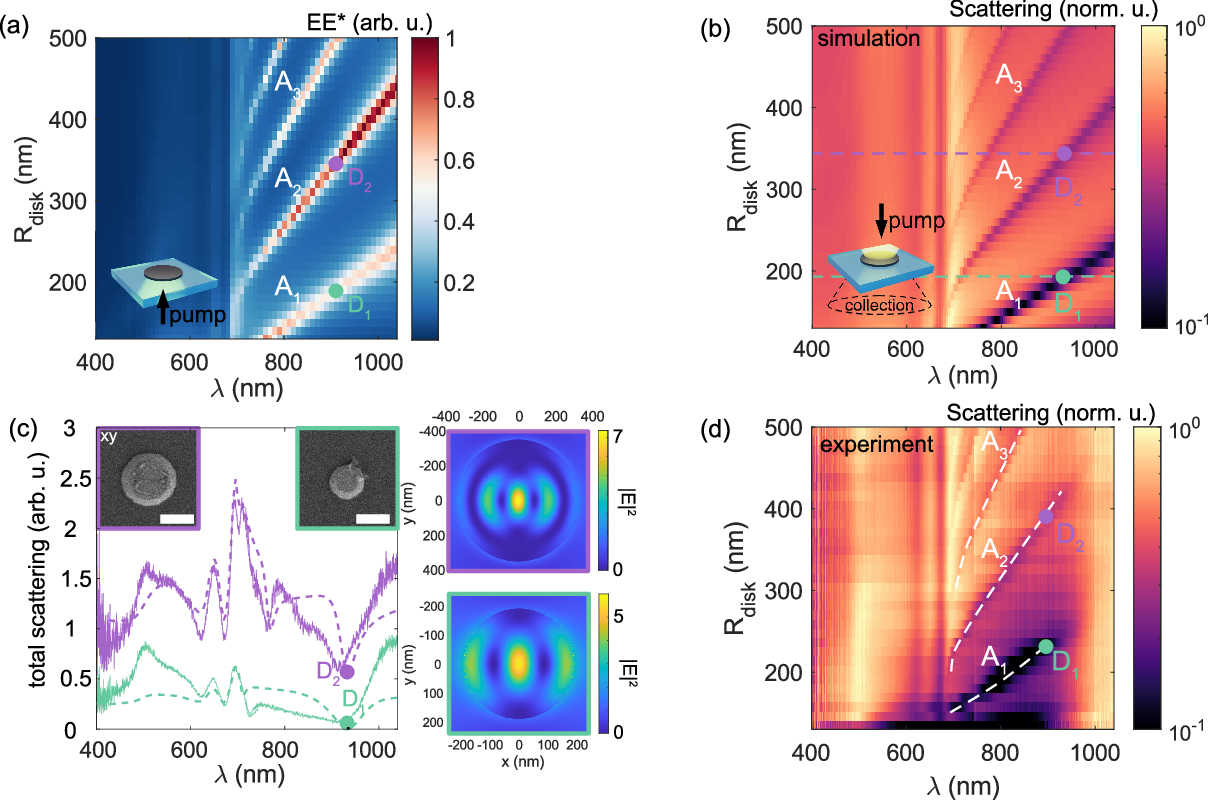}
    \caption{\textbf{Linear optical properties of 3R-\ce{MoS2} nanodisks.} a) Numerical calculation of $|E|^2$, characterizing the stored electromagnetic energy inside the 3R-\ce{MoS2} nanodisk, in normalized units. A$_1$, A$_2$, A$_3$ -- anapole-like states. Inset shows the schematics of the calculation geometry. D$_1$ and D$_2$ label nanodisks with resonantly enhanced stored electromagnetic energy. b) Numerical calculation of the elastic light scattering spectra by individual 3R-\ce{MoS2} nanodisks as a function of the nanodisk's radius in dark-field geometry using conditions similar to experimental ones. c) Lilac ($R_{\mathrm{disk}}$ = 350~nm) and mint (190~nm) lines correspond to D$_2$ and D$_1$ numerical (dashed) and experimental (solid) dark-field scatterings. Inset shows SEM images of the two selected nanodisks. The scale bar is 500~nm. Side panels show the corresponding near-field distributions calculated at a 920~nm pump. d) The color map shows experimental dark-field scattering spectra for 3R-\ce{MoS2} nanodisks in a logarithmic scale. Each scattering spectrum obtained for a particular radius $R_\mathrm{disk}$ is normalized to its maximum value. The white dashed lines indicate the dispersion of anapole states with the nanodisk's radius.}
    \label{fig:2}
\end{figure}

We start with a numerical design of an appropriate nanodisk geometry. Our analysis involves studying resonantly enhanced electromagnetic energy stored within the volume of the nanodisk, namely $u_{\mathrm{NP}}\propto |E|^2$, where $|E|^2 = EE^*$, and $E^*$ stands for complex conjugation of the electric field. \autoref{fig:2}a shows the colormap of normalized $|E|^2$ in the volume of an individual 3R-\ce{MoS2} nanodisk of 65~nm height and a range of radii $R_{\mathrm{disk}}$ from 130~nm to 500~nm under normal plane-wave excitation from the bottom (glass substrate) side. Such excitation geometry is chosen because it resembles the experimental SHG setup (discussed below), which allows for improved scattering signal collection due to a higher $k$-vector in the glass substrate. The choice of a 3R-\ce{MoS2} nanodisk height of $h = 65$~nm is based on our numerical analysis, which indicates that the optimal SHG response occurs within the range of heights $h \approx$ 60 -- 90~nm for a free-standing nanodisk.
Furthermore, our numerical analysis allows for using directional excitation and specific apertures for collecting the scattered light. One can see that the nanodisks host three resonant features (denoted as A$_1$, A$_2$, A$_3$) that show enhanced $|E|^2$ in the spectral region of the highest nonlinear response. The region below 700~nm demonstrates no resonant features due to high optical losses. We mark the optimal disks with points D$_1$ and D$_2$, corresponding to the highest stored electromagnetic energy at the wavelength of the material $\chi^{(2)}$ resonance ($\sim$ 920~nm). These disks are promising for subsequent SHG experiments and their theoretically predicted parameters are $R_{\mathrm{disk}}$ = 190~nm and 350~nm, respectively.

In order to assign the resonant features A$_1$, A$_2$, A$_3$ to optical anapoles, we first calculate the scattering response of the nanodisks. \autoref{fig:2}b shows a colormap of the calculated scattering cross-section for various 3R-\ce{MoS2} nanodisk sizes. Note that in this case, the light is incident on the sample from the top, to mimic the experimental dark-field (DF) scattering setup, discussed below. The scattering cross-section at each specific nanodisk radius is normalized to unity to enhance the visual representation of the data. Indeed, the A$_1$, A$_2$, and A$_3$ features demonstrate anapole-like behavior, which is manifested in a significant drop in their scattering signal. The scattering spectra of D$_1$ and D$_2$ nanodisks (mint and lilac color dashed lines) alongside their calculated electromagnetic near-field maps are shown in \autoref{fig:2}c. The near-field maps were evaluated at the minima of the scattering spectra (labeled as D$_1$ and D$_2$) and show that the electric fields are confined mostly within the nanodisk volume and exhibit typical anapole profiles. We thus conclude that within the studied parameter range, the appearance of optical anapoles in 3R-\ce{MoS2} nanodisks is feasible \cite{verre2019transition}.

Based on the theoretical predictions, we fabricated 65~nm high 3R-\ce{MoS2} nanodisks with radii in the range of 130 -- 500~nm. Such a design results in a number of resonant features in the 880 -- 960~nm spectral range, making it highly promising for simultaneous enhancement of SHG through the combined effect of optical and material resonances.

We performed morphological characterization of the fabricated nanodisks
by scanning electron microscopy (SEM) and atomic-force microscopy.
\autoref{fig:2}c depicts experimental DF scattering of two individual nanodisks with experimental radii of $\sim$ 225~nm (mint solid) and $\sim$ 380~nm (lilac solid). These spectra are plotted against the theoretical 190~nm and 350~nm radii disks (corresponding colors -- dashed line). The agreement between theory and experiment is good, with an important remark that there is a slight offset between the sizes used in the calculation model and the actual experimental sizes. This difference may arise due to the disk's geometry deviating from a perfect cylinder and having a slight frustum shape (bottom and top radii are slightly different). The insets in \autoref{fig:2}c depict the corresponding SEM images of experimentally fabricated nanodisks, demonstrating this point. Taking this into account, the experimental DF scattering response is slightly shifted in comparison to the numerical design, as one can see from \autoref{fig:2}d.
In \autoref{fig:2}d, we highlight the anapole-like dips A$_{1-3}$ observed in experimental DF scattering spectra and their dispersion with the size of the nanodisks. 

\begin{figure}
    \centering
    \includegraphics[width=0.7\linewidth]{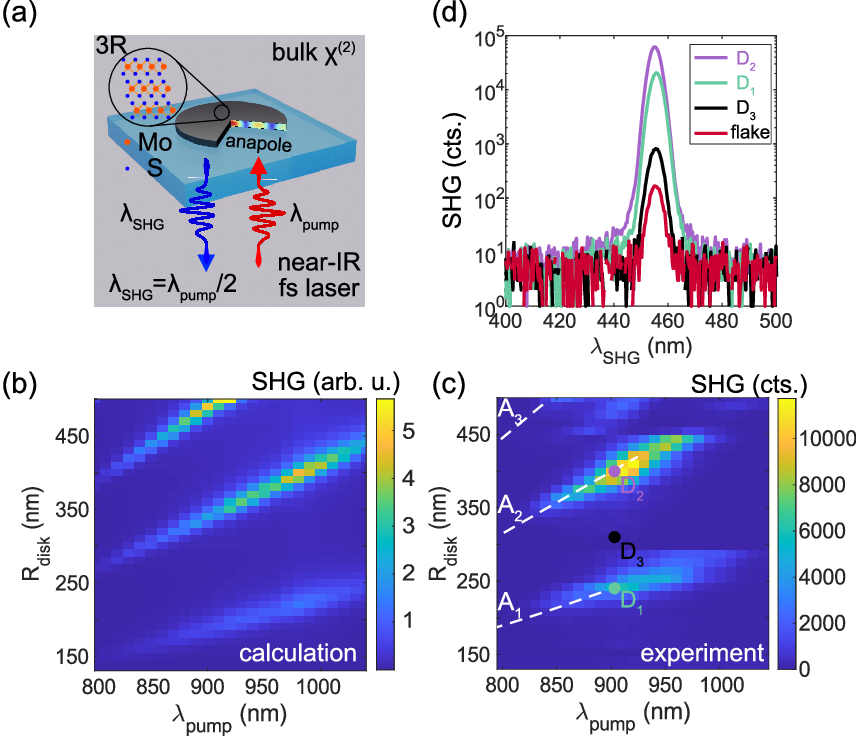}
    \caption{\textbf{Second-harmonic generation in 3R-\ce{MoS2} nanodisks.} a) Schematic of the experiment: an individual 3R-\ce{MoS2} nanodisk is excited with a near-infrared femtosecond laser and efficiently converts the pump light into SHG due to a combination of the anapole state hosted by the nanodisk and a substantial $\chi^{(2)}$ nonlinearity. A quarter of the nanodisk has been removed to visualize the near-field distribution at the anapole state.  b) Numerical calculation of the SHG intensity of 3R-\ce{MoS2} nanodisks with varying disk radii on a glass substrate at different pump wavelengths. The SHG intensity is obtained as total radiation generated at  $\lambda_{\mathrm{SHG}}$=$\lambda_{\mathrm{pump}}/2$. c) Experimentally measured SHG color map of individual 3R-\ce{MoS2} nanodisks on a glass substrate at different pump wavelengths. The intensity of the SHG signal is obtained at $\lambda_{\mathrm{pump}}/2$ for every nanodisk. d) SHG spectra measured at 910~nm pump for an unpatterned 3R-\ce{MoS2} flake (red line), resonant nanodisks: D$_1$ (mint) and D$_2$ (lilac), and an arbitrary off-resonant nanodisk D$_3$ (black). Note that the $y$-axis is shown in a logarithmic scale.}
    \label{fig:3}
\end{figure}

\textbf{Resonantly-enhanced SHG in individual 3R-\ce{MoS2} nanodisks.}
We now turn our attention to SHG of individual 3R-\ce{MoS2} nanodisks. Specifically, we study their nonlinear optical properties in a configuration schematically shown in \autoref{fig:3}a. The SHG intensity calculation for nanodisks adopting this excitation-collection geometry is shown in \autoref{fig:3}b. A Comsol Multiphysics numerical model was used in two steps: i) the near-infrared wavelength excitation $\lambda_{\mathrm{pump}}$ of the nanodisk followed by an extraction of the near-field data; ii) evaluation of the nonlinear polarization of the disk at $\lambda_{\mathrm{pump}}/2$ using near-fields from i) weighted by the $\chi^{(2)}$ tensor, and, finally, SHG at the $\lambda_{\mathrm{pump}}/2$. Remarkably, the areas of the highest SHG intensity in the spectra appear at the overlap between the $\chi^{(2)}$ tensor resonance (\autoref{fig:1}b) and in good agreement with the A$_1$, A$_2$, and A$_3$ curves predicted in the linear response (\autoref{fig:2}a).

We further proceed to SHG measurements. \autoref{fig:3}c shows SHG maps obtained for nanodisk radii in the 130 -- 500~nm range and pump wavelengths 800 -- 1040~nm (fixed power of $\sim$ 300~$\mu$W). The data was recorded by scanning the sample with respect to the focused laser spot over the entire nanodisk array (5 $\mu$m disk-to-disk separation) at a given pump wavelength using a piezo stage. The nonlinear signal was collected using an APD. 
As evaluated by the stored electromagnetic energy and linear DF scattering analysis (\autoref{fig:2}), the nanodisks with experimental radii of $\sim$ 225~nm (D$_1$) and $\sim$ 380~nm (D$_2$) support anapole-like behavior at the wavelength at which we observe maximum experimental $\chi^{(2)}$ (see \autoref{fig:1}). These disks, therefore, are expected to display the highest SHG intensity peaking at $\sim$ 910 -- 920~nm $\lambda_{\mathrm{pump}}$, which matches the condition of the highest experimental value of $\chi^{(2)}$ for the material. Remarkably, the experimentally measured anapole-enhanced SHG for a 380~nm nanodisk (D$_2$) at a 910~nm pump provides close to 2 orders ($\sim$ 80-fold) of magnitude increase in the SHG signal compared to the arbitrary off-resonant nanodisk at this wavelength (labeled as D$_3$ with $R_{\mathrm{disk}}~\sim$ 310 nm).
Moreover, the difference between the simultaneously optically and materially resonant D$_2$ nanodisk pumped at $\lambda_{\mathrm{pump}}= 910~\mathrm{nm~}$ 
and an off-resonant (both optically and materially) one can be significantly higher and approaches $\sim4$ orders of magnitude (limited by detection resolution, $\sim$ 5000-fold). 
An SHG enhancement exceeding 2-orders of magnitude  is reached between the resonant anapole nanodisk and the parent flake from which the disk was fabricated. However, in the case of the unpatterned flake, the volume of the emitting SHG material is likely much larger due to the diffraction limit of the pump beam. Remarkably, the highest SHG intensity areas spectrally match with the DF scattering anapole dips confirming the hypothesis that the SHG signal is dramatically enhanced by optical anapole states. Furthermore, the agreement with the calculations of the SHG response of the nanodisks on a glass substrate shown in \autoref{fig:3}b is also remarkable. A slight mismatch between calculations and experiments is mainly due to two reasons: i) the DFT-calculated $\chi^{(2)}$ tensor, used for numerical calculations, peaks around 950 -- 960~nm pump wavelength, whereas the experimental one is around 910 -- 920~nm (\autoref{fig:1}b), and, ii) the SEM images demonstrate that the fabricated nanodisks slightly deviate from the perfect cylindrical shape, which affects their optical response.

To further improve our understanding of the SHG enhancement mechanism, we have performed an explicit head-to-head comparison of the SHG signals from the two most resonant disks D$_1$, D$_2$, the non-resonant disk D$_3$ (at the same $\lambda_\mathrm{pump}$ = 910~nm wavelength, but without anapole enhancement) and the unpatterned parent flake. Additionally, we collected the SHG signal of these disks under off-material resonance conditions under 820 and \unit[1020]{nm} pumping. These measurements show SHG signals slightly above the background noise level. 
Such a comparison allows us to roughly estimate the relative contributions of material and optical enhancements to the SHG signal. Specifically, our analysis reveals that the optical anapole enhancement is responsible for about 80-fold enhancement (from the comparison between D$_1$, D$_2$, and D$_3$). The material resonance provides an additional 100-fold increase at the 910 nm pump wavelength when compared to 820 to 1020 nm pumps (from \autoref{fig:1}b). Jointly, the optical and material resonance grant SHG signal tuning in an enhancement range approaches 4 orders of magnitude within the near-IR spectral range ($\sim$ 5000-fold).
This is illustrated in \autoref{fig:3}d, which depicts the SHG spectra (log scale) obtained using a 910~nm pump wavelength at low pump power \unit[0.3]{mW} and just \unit[1]{s} collection time. The two most resonant disks -- D$_2$ (lilac line) and D$_1$ (mint line) possess an SHG signal that is more than 2 orders of magnitude stronger than the unpatterned 3R-\ce{MoS2} parent flake (red), whereas the off-resonant nanodisk D$_3$ is only slightly stronger than the unpatterned flake. 
One of the most off-resonant disks in which an SHG signal can be detected with the spectrometer at the same experimental conditions is the D$_1$ disk pumped at \unit[820]{nm}. As a consequence of being driven outside the anapole state and at the edge of the $\chi^{(2)}$ resonance, its SHG spectrum at \unit[410]{nm} peaks slightly above the noise floor and is approximately 1 and 2 orders of magnitude weaker than, respectively, both the unpatterned flake and optically off-resonant disk D$_3$ when driven at the material resonance (see \autoref{fig:3}d).

In conclusion, we demonstrated that the SHG emission can be dramatically enhanced in bulk $\chi^{(2)}$ 3R-\ce{MoS2} material via combining material and nanophotonic resonances. The material response was optimized by exciting the resonance of the $\chi^{(2)}$ at $\sim$ 910~nm pump wavelength, whereas nanophotonic resonances were engineered through appropriate nanodisk dimensions supporting the anapole state in the wavelength range overlapping with the material resonance. Such an approach provides more than 3 orders of magnitude enhancement of SHG signal for the most resonant nanodisks \emph{vs.} the most off-resonant case ($\sim$ 5000-fold) and more than 2 orders of magnitude enhancement compared to the unpatterned flake of the same thickness ($\sim$ 400-fold). We anticipate that by manipulating the thickness of the 3R-\ce{MoS2} flake, one can potentially achieve even higher enhancement of SHG by engineering nanophotonic resonances. Moreover, we envision that nanostructured 3R-\ce{MoS2} holds great potential as a versatile platform for a wide range of all-TMD nonlinear nanophotonics applications beyond SHG. This includes various $\chi^{(2)}$ nonlinear processes such as the generation of entangled photon pairs \cite{santiago2022resonant}, optical parametric amplification, electro-optical effects, and more, wherein the resonator and nonlinear medium are integrated within the same nanostructured object, resonant metasurface, or nanophotonic circuit.

\section*{Acknowledgments}

G.Z., A.P., B.K., and T.O.S acknowledge funding from the Swedish Research Council (VR Miljö project, grant No: 2016-06059 and VR project, grant No: 2017-04545), the Knut and Alice Wallenberg Foundation (grant No: 2019.0140), Chalmers Area of Advance Nano, 2D-TECH VINNOVA competence center (Ref. 2019-00068) and Olle Engkvist foundation (grant No: 211-0063). M.B. and T.J.A. acknowledge support from the Polish National Science Center via the project 2019/34/E/ST3/00359. This work was performed in part at Myfab Chalmers and at the Chalmers Material Analysis Laboratory, CMAL. Calculations were partially done at the ICM, UW (\#GC84-51).



\section*{Conflicts of interests}
The authors declare no conflicts of interest.



\newpage


\end{document}